# A Computational Study of Yttria-Stabilized Zirconia: II. Cation Diffusion

Yanhao Dong[1], Liang Qi[2], Ju Li[3] and I-Wei Chen[1*]

[1]Department of Materials Science and Engineering, University of Pennsylvania,

Philadelphia, PA 19104, USA

[2]Department of Materials Science and Engineering, University of Michigan, Ann

Arbor, MI 48109, USA

[3]Department of Nuclear Science and Engineering and Department of Materials

Science and Engineering, Massachusetts Institute of Technology, Cambridge, MA

02139, USA

**Abstract**

Cubic yttria-stabilized zirconia is widely used in industrial electrochemical devices. While its fast oxygen ion diffusion is well understood, why cation diffusion is much slower—its activation energy (~5 eV) is 10 times that of anion diffusion—remains a mystery. Indeed, all previous computational studies predicted more than 5 eV is needed for forming a cation defect, and another 5 eV for moving one. In contrast, our *ab initio* calculations have correctly predicted the experimentally observed cation diffusivity. We found Schottky pairs are the dominant defects that provide cation vacancies, and their local environments and migrating path are dictated by packing preferences. As a cation exchanges position with a neighboring vacancy, it passes by an empty interstitial site and severely displaces two oxygen neighbors with shortened Zr-O distances. This causes a short-range repulsion against the migrating cation and a long-range disturbance of the surrounding, which explains why cation diffusion is relatively difficult. In comparison, cubic zirconia's migrating oxygen only minimally disturbs neighboring Zr, which explains why it is a fast oxygen conductor.

[*]Corresponding Author
*E-mail address*: iweichen@seas.upenn.edu (I-Wei Chen)

I. Introduction

Zirconia ceramics have important applications such as structural components, thermal barrier coating, solid electrolytes and gas sensors. They are used in the tetragonal or cubic form, stabilized by aliovalent cations such as $Y^{3+}$ and $Ca^{2+}$. These cations are compensated by oxygen vacancies that make stabilized zirconia a good $O^{2-}$ conductor. However, cation transport at 1,000 ºC is more than one-trillion-fold slower than $O^{2-}$ [1], and it is the diffusion of cations being the slowest-moving species that determines the total mass flow, which is central to microstructural control in processing (e.g., sintering and grain growth) and high temperature service (e.g., in thermal barrier coating and fuel cell). In cubic yttria-stabilized zirconia (YSZ), cation lattice diffusivity inferred from dislocation loop shrinkage [2], creep [3-4] and tracer migration [1, 5] is $\sim 10^{-14}$ cm$^2$/s at 1,500 ºC with apparent activation energy of 4.5-6.1 eV. Such activation energy is consistent with the common experience that zirconia sintering is usually performed at 1,300-1,550 ºC [6, 7]. It is also consistent with grain growth kinetics in tetragonal zirconia (2 mol% yttria-stabilized zirconia, 2YSZ) [7], which has activation energy of 4.6 eV that is attributed to cation lattice diffusion since 2YSZ's grain boundary mobility is controlled by solute-cation drag in the lattice [8]. On the other hand, all previous theoretical and computational studies have consistently found an activation energy >10 eV, of which 6.5 eV or more is for forming a cation defect [9-12] and 5 eV for the defect to migrate [11-13]. Such activation energy is undoubtedly too high as it will rule out any kinetic possibility at all temperatures up to zirconia's melting point, 2,750 °C. (To reach a diffusion distance of 1 μm after 1,000 s or $10^{-17}$ m$^2$/s in diffusivity at temperature $T$, the activation energy cannot exceed $25k_BT$.) Thus, a much better computational study and understanding is needed to align with the experimental observations.

This study will address the above need. As we already noted in the companion paper (hereafter referred to as Paper I) [14], all the previous computational studies on cation defect and diffusion in YSZ employed empirical potential calculations [9-13, 15]. This is not surprising because YSZ contains not only two types of cations but also many anion vacancies, which generate an astronomically large number of configurations even for a small supercell. In Paper I, we have conducted *ab initio* and empirical-potential calculations for a number of supercell configurations, and found the latter always overestimate the energy by a factor of two or so primarily because they grossly overestimate the electrostatic energy. Therefore, this study will only use *ab initio* calculations to compute defect's formation and migration energies.

These computations will mainly be performed with the "ground" state of YSZ (established in Paper I), in which we shall find activated cation defects—e.g., Schottky pairs and cation Frenkel pairs—and study their motion. We will also assess the diffusion contribution of metastable states in the glassy energy landscape of YSZ (see Paper I).

The paper is organized as follows. After describing the simulation and calculation methods in Section II, we calculate in Section III the formation energies of a Schottky pair, a cation Frenkel pair and an anion Frenkel pair for three forms of zirconia to provide a reference frame and preliminary insight. In Section IV, the formation energies of a Schottky and a cation Frenkel pair in the YSZ ground state and some metastable states are calculated to ascertain cation vacancies as the dominant cation defect species. In Section V, their migration barriers are calculated for Zr and Y, some with nearby oxygen vacancies. In Section VI, the migration barrier for serial random-walk events is computed to compare with experimental diffusivity. This is followed by discussions in Section VII and conclusions in Section VIII. The work is focused on 8 mol% yttria stabilized zirconia (8YSZ), which has attracted much theoretical and practical interest.

## II. Methodology

We used the projector augmented-wave (PAW) method [16] and the Perdew-Burke-Ernzerhof (PBE) [17] generalized gradient approximation (GGA) implemented in the Vienna *ab initio* simulation package (VASP) [18]. The PAW potentials include the following electrons: $5s^24d^2$ for Zr, $4s^24p^65s^24d^1$ for Y and $2s^22p^4$ for O. We chose a plane-wave cutoff energy of 500 eV to reach a convergence criterion of 1 meV for the total energy and sampled the Brillouin zone using the Monhorst-Pack scheme with a 2×2×2 $k$-point mesh. These results will be referred to as *ab initio* calculated data in the following. Where appropriate, empirical potential calculations [19] using General Utility Lattice Program (GULP) [20], as described in Paper I, was too performed. For ZrO2 in its monoclinic, tetragonal or cubic form, a supercell containing 108 Zr and 216 O was used, corresponding to a 3×3×3 supercell for the monoclinic and cubic phase and a 3×3×6 supercell for the tetragonal phase. For YSZ, a similar 3×3×3 supercell with 92 Zr, 16 Y and 208 O was used, simulating 8YSZ of the same stoichiometry. When the supercell contains a point defect, such defect is assigned a formal charge with respect to the reference state according to the Kröger-Vink notation (-4 for Zr vacancy $V_{Zr}$, +4 for Zr interstitial $Zr_i$, -3 for Y vacancy $V_Y$, +3 for Y interstitial $Y_i$, and +2 for oxygen vacancy $V_O$; all in electron unit). An opposite charge was applied as a uniform background to ensure (a) cations and anion are not reduced or oxidized and (b) the supercell remains neutral. For defect migration in YSZ, the nudged-elastic-band (NEB) method [21] implemented in VASP was used to determine the diffusion path and migration barrier under a fixed supercell volume and shape. Here, we used the same 3×3×3 supercell, and convergence was considered achieved when the residue atomic forces are less than 0.1 eV/Å. To help accurately evaluate the saddle point energy, we employed 7 climbing images between the initial and final configurations. All calculations were performed under periodic boundary conditions.

**III. Defect formation energy in pure zirconia**

We first describe the formation energies of a Schottky pair and a cation/anion Frenkel pair in pure $ZrO_2$ of the monoclinic, tetragonal and cubic structures as shown in **Fig. 1**. For monoclinic $ZrO_2$ (space group P 21/c), Zr and O occupy different 4*e* sites and interstitials are at 2*c* sites at (0 1/2 0) and (0 0 1/2). For tetragonal $ZrO_2$ (space group P 42/nmc), O occupies 4*d* sites, Zr at 2*a* sites at (0 0 0) and (1/2 1/2 1/2) and interstitials at 2*b* sites at (0 0 1/2) and (1/2 1/2 0). (Alternative arrangements may be made by interchanging 2*a* and 2*b* sites, which are related by a (1/2 1/2 0) translation.) For cubic $ZrO_2$ (ideal fluorite structure, space group $Fm\bar{3}m$), O occupies 8*c* sites at (1/4 1/4 1/4) and (1/4 1/4 3/4) adopting simple cubic packing, while Zr occupies 4*a* sites at (0 0 0) and interstitials at 4*b* sites at (1/2 1/2 1/2), both adopting a face center cubic packing. (Alternatively, Zr can occupy 4*b* sites while the interstitials occupy 4*a* sites, which differs from the above arrangement by a (1/2 1/2 1/2) translation.) Since all lattice vacancy/interstitial locations are equivalent, any of them may be chosen for calculating defect formation energies after allowing for relaxations of supercell's shape, volume and atomic positions.

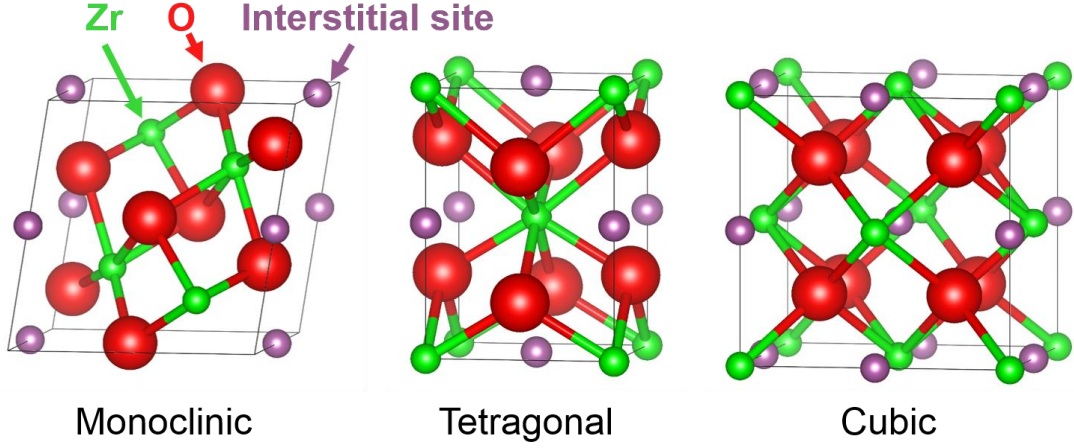

Figure 1 Crystal structures of monoclinic, tetragonal and cubic $ZrO_2$. The unit cell contains 4 Zr and 8 O for monoclinic and cubic $ZrO_2$, and 2 Zr and 4 O for tetragonal $ZrO_2$. Symbols: Zr in green, O in red and interstitial sites in purple.

In the Kröger-Vink notation, the defect reactions to form a cation Frenkel pair and a Schottky pair and are, respectively.

$$Zr_{Zr}^{\times} = V_{Zr}''''+ Zr_i^{\bullet\bullet\bullet\bullet} \quad (1)$$

$$Nil = V_{Zr}''''+ 2V_O^{\bullet\bullet} \quad (2)$$

Thus, the energy to form a cation Frenkel pair, $E_{f,\text{ cation Frenkel}}$, is

$$E_{f,\text{ cation Frenkel}} = E_{V_{Zr}} + E_{Zr_i} - 2E_{ref} \quad (3)$$

where $E_{V_{Zr}}$ denotes the energy of a supercell that containing one $V_{Zr}$, and likewise $E_{Zr_i}$ is the supercell energy with one $Zr_i$, whereas $E_{ref}$ is the energy of a reference $ZrO_2$ supercell without any defect. The energy to form an anion Frenkel pair is similarly obtained. To form a Schottky pair, $E_{f,\text{ Schottky}}$, we also need to consider the energy of (a) removing one Zr and two O, and (b) depositing them in a reservoir that holds the chemical potential of $ZrO_2$, which is set as $\mu_{ZrO_2} = E_{ref}/108$ by referring to the defect-free supercell where there are 108 pairs of $ZrO_2$. Thus,

$$E_{f,\text{Schottky}} = E_{V_{Zr}} + 2E_{V_O} - 3E_{\text{ref}} + N \cdot \mu_{ZrO_2} = E_{V_{Zr}} + 2E_{V_O} - \frac{323}{108}E_{\text{ref}} \quad (4)$$

where $N=1$ and $E_{V_O}$ is the energy of a supercell that contains one $V_O$.

For each $ZrO_2$ polymorph in a 3×3×3 supercell, we performed *ab initio* calculations to find the supercell energies with and without various defects, then used Eq. (3-4) to obtain the pair formation energies listed in **Table I**. Without exception, $E_{f,\text{Schottky}}$ is less than $E_{f,\text{cation Frenkel}}$, and the difference increases from monoclinic to tetragonal to cubic. This is mainly due to a corresponding decrease of $E_{f,\text{Schottky}}$ since $E_{f,\text{cation Frenkel}}$ is almost the same in all three polymorphs. Interestingly, the much more open monoclinic structure (its unit cell volume is 4% larger) that should help accommodate an interstitial does not make $E_{f,\text{cation Frenkel}}$ smaller. The different $E_{f,\text{Schottky}}$ may stem from two related causes: (a) As the structure changes from monoclinic to cubic, $Zr^{4+}$ is coordinated with more oxygens, which is unfavorable because of the small size of $Zr^{4+}$, and (b) as Zr becomes overbonded in the tetragonal and cubic $ZrO_2$, the Zr-O bond strength weakens. Therefore, oxygen removal and $V_O$ formation, which plays a dominant part in a Schottky pair that contains two $V_O$, is easiest in the cubic structure and most difficult in the monoclinic structure. In this connection, it is interesting to note that, in cubic zirconia, the Schottky pair reaction creating three defects—one $V_{Zr}$ and two $V_O$—provides the lowest formation energy per defect, $E_{f,\text{Schottky}}/3 = 0.91\text{ eV}$; the anion Frenkel pair reaction (creating two defects—one $V_O$ and one $O_i$) gives $E_{f,\text{anion Frenkel}}/2 = 1.54\text{ eV}$. Thus, in an undoped cubic $ZrO_2$, $V_O$ is mainly created by the Schottky pair reaction rather than the anion Frenkel pair reaction.

**Table I** Formation energies of Schottky pair, cation Frenkel pair and anion Frenkel pair in $ZrO_2$.

| Crystal type | Schottky pair (eV) | Cation Frenkel pair (eV) | Anion Frenkel pair (eV) |
|---|---|---|---|
| Monoclinic | 7.06 | 8.75 | 4.26 |
| Tetragonal | 3.98 | 8.14 | 3.45 |
| Cubic | 2.84 | 8.56 | 3.09 |

Since a Schottky pair requires less energy to form in all three polymorphs, $V_{Zr}$ is the main cation defect species, especially in tetragonal and cubic zirconia. While this conclusion agrees with the previous computational studies [9-11], our $E_{f, Schottky}$ for cubic zirconia (2.84 eV) is much smaller than the ones previously calculated (11.6 eV by Mackrodt *et al.* [9] and 6.52 eV by Kilo *et al.* [11]). Similarly, our $E_{f, cation Frenkel}$ (8.56 eV) is much smaller (24.4 eV by Mackrodt *et al.* [9] and 20.4 eV by Kilo *et al.* [11]). Since there is no structural ambiguity about cubic $ZrO_2$, the difference can only come from the computational method itself. This trend parallels a similar one established in Paper I: Empirical potential calculations consistently produced >2× supercell energies of what the *ab initio* calculations produced. The cause of this error is also known: It is primarily due to an overestimate of the electrostatic interaction [14].

## IV. Defect formation energy in YSZ

4.1 Formulation of the problem

Unlike in ZrO2, defect pairs in YSZ include partial defect pairs of different constituents. The defect reactions forming a Schottky pair and a cation Frenkel pair in $Zr_{1-x}Y_xO_{2-x/2}$ are, respectively,

$$\text{Nil} = (1-x)V_{Zr}'''' + xV_Y''' + (2-\frac{x}{2})V_O^{\bullet\bullet} \quad (5)$$

$$(1-x)Zr_{Zr}^{\times} + xY_Y^{\times} = (1-x)V_{Zr}'''' + (1-x)Zr_i^{\bullet\bullet\bullet\bullet} + xV_Y''' + xY_i^{\bullet\bullet\bullet} \quad (6)$$

where the reference cation sublattice includes both Zr's and Y's in their stoichiometric proportion. Referring to the energies of supercells that contain various point defects in the above reactions, we can express the formation energies of a Schottky pair and a cation Frenkel pair as

$$\begin{aligned}E_{f,\text{Schottky}} &= (1-x)E_{V_{Zr}} + xE_{V_Y} + (2-\frac{x}{2})E_{V_O} - (3-\frac{x}{2})E_{ref} + N \cdot \mu_{YSZ} \\ &= (1-x)E_{V_{Zr}} + xE_{V_Y} + (2-\frac{x}{2})E_{V_O} - (\frac{323}{108} - \frac{x}{2})E_{ref}\end{aligned} \quad (7)$$

$$E_{f,\text{cation Frenkel}} = (1-x)E_{V_{Zr}} + (1-x)E_{Zr_i} + xE_{V_Y} + xE_{Y_i} - 2E_{ref} \quad (8)$$

In the above, $E_{V_Y}$ denotes the energy of a supercell that contains one $V_Y$, and likewise $E_{Y_i}$ is the energy of a supercell of one $Y_i$, etc. Here, $E_{ref}$ is the energy of a reference defect-free supercell of $Zr_{1-x}Y_xO_{2-x/2}$, $N=1$ refers to one set of $Zr_{1-x}Y_xO_{2-x/2}$ "molecule", and $\mu_{YSZ} = E_{ref}/108$ is the chemical potential of such "molecule". To apply it to 8YSZ, we let $x=16/108\approx0.148$.

To compute $E_{V_Y}$, etc., we must recognize that unlike in ZrO2, there is no translational invariance in YSZ; e.g., the local environments of any two ZrZr are generally different, etc. So, starting with a configuration of defect-free YSZ, there are many distinct choices for placing a point defect, say $Y_i$, and each choice gives a different supercell energy $E_{Y_i}$. In a 3×3×3 supercell of 8YSZ, there are 108 choices each for placing a $Zr_i$ or $Y_i$, 92 choices for placing a $V_{Zr}$, 16 for $V_Y$, and 208 for $V_O$. This gives 532 in total, which is too many to be handled by *ab initio* calculations. To keep the computational task tractable, we first examined all these choices by empirical potential calculations and ranked them in the order of the lowest supercell energy. We next selected a certain number of the lowest energy configurations, 10 each for $V_{Zr}$ and $Zr_i$, 5 each for $V_Y$ and $Y_i$, and 20 for $V_O$, and computed their supercell energies using *ab initio* calculations to obtain $E_{V_Y}$, etc. They do show good correlation with the values obtained by empirical potential calculations, as shown in **Fig. 2**, with the magnitude of the latter more than twice larger, which was also noted in Paper I when the two calculations were compared. Substituting thus *ab initio* calculated $E_{V_{Zr}}$, $E_{Zr_i}$, $E_{V_Y}$, $E_{Y_i}$ and $E_{V_O}$ in various combinations into Eq. (7-8), we obtained a distribution of $E_{f,\,Schottky}$ and $E_{f,\,cation\,Frenkel}$, which spans certain ranges reflecting the combinations of different point defects at different sites. In Paper I, a similar hybrid strategy combining empirical potential calculations for screening and *ab initio* calculations for final results was used to assess the stability of supercells. Although the two calculations returned rather different values for the supercell energies, the two sets of energies do track with each other with a scaling factor of 2 or so (see **Fig. 5** in Paper I). So we believe it is highly likely that the above strategy will capture the least energetic Schottky and cation Frenkel pairs: These pairs is likely to be represented in the low-energy-tail of the above distributions. Specifically, the least energetic Schottky pair must consist of the least energetic $V_{Zr}$, $V_Y$ and $V_O$, and the least energetic cation Frenkel pair must consist of the least energetic $V_{Zr}$, $V_Y$, $Zr_i$ and $Y_i$. Another way to state the same is that the least energetic Schottky pair must come from removing the most energetic $Zr_{Zr}$, $Y_Y$ and $O_O$, etc.

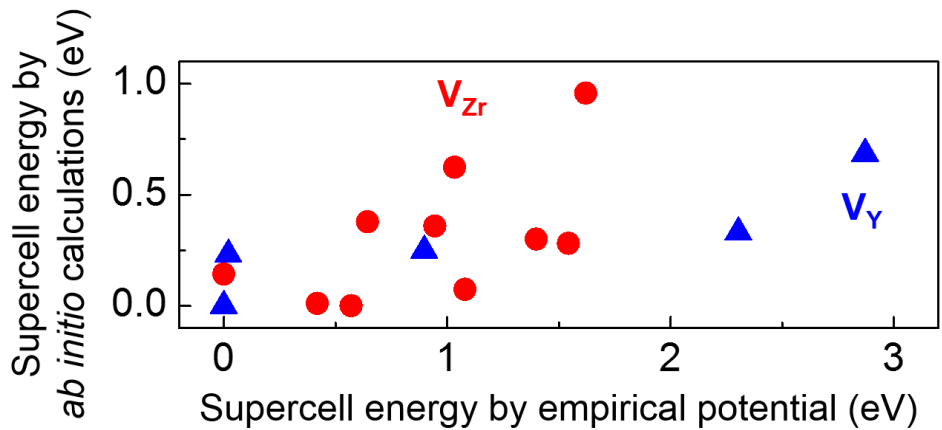

Figure 2 Comparison of supercell energies containing one $V_{Zr}$ (in red) or one $V_Y$ (in blue) by *ab initio* and empirical potential calculations. The lowest energy in each set is set as zero.

4.2 Schottky pairs and cation Frenkel pairs in the ground state

Starting with the "ground" state configuration (**Fig. 3a**) and its energy (which gives $E_{ref}$ and $\mu_{YSZ}$) identified in Paper I, we obtained a set of distributions of formation energies shown in **Fig. 3b-c.** The large spread of the pair energies comes from the energy spreads of $E$'s of individual defects in the pair. For example, since the 10 lowest $E_{V_{Zr}}$ spread over 0.96 eV, and likewise 5 $E_{V_Y}$ over 0.69 eV and 20 $E_{V_O}$ over 1.55 eV, the spread of $E_{f, Schottky}$ is $(1-x) \times 0.96 + x \times 0.69 + (2 - \frac{x}{2}) \times 1.55 = 3.90$ eV. Applying Boltzmann statistics to these distributions over the temperature of 1400-1700K, we calculated the effective formation energies to be 1.91 eV for a Schottky pair and 5.77 eV for a cation Frenkel pair (**Fig. 3d**). Therefore, as in pure cubic $ZrO_2$, a Schottky pair in YSZ is much easier to form, making cation vacancies the dominant cation defects and Schottky pair reaction their dominant source. (Using the law of mass action, the concentration of cation interstitials is estimated to be 9-10 orders smaller than that of cation vacancies.) Lastly, since Y doping introduces a large population of $V_O$, the concentration and chemical potential of $V_O$ in YSZ are fixed (i.e., YSZ is in the extrinsic regime). So the concentration of cation vacancy can be obtained from the law of mass action for Eq. (2), the Schottky defect reaction with an reaction energy $E_{f,\text{Schottky}}$, giving $[V_{Zr}] \sim \dfrac{1}{[V_O^{\bullet\bullet}]^2} \exp\left(-\dfrac{E_{f,\text{Schottky}}}{k_B T}\right)$.

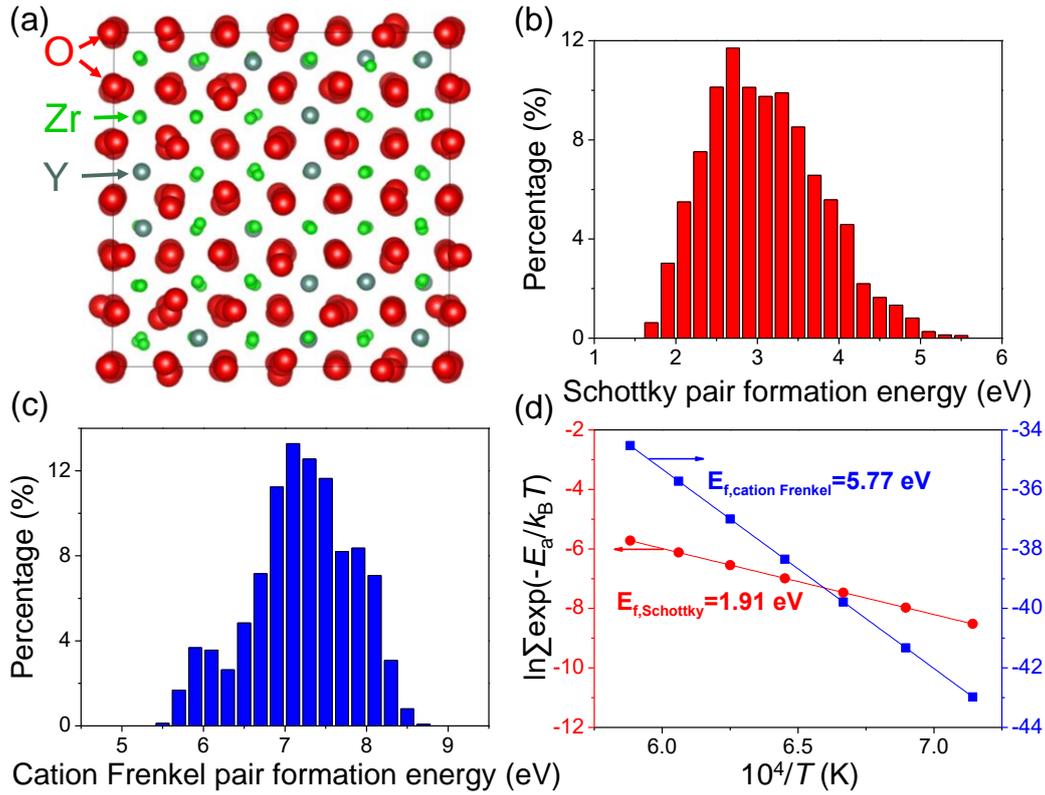

**Figure 3** (a) "Ground" state YSZ configuration identified in Paper I and the energy distributions to form (b) a Schottky pair or (c) a cation Frenkel pair in the structure. (d) The assemble averages of defect formation energy between 1400-1700 K. Data from *ab initio* calculations.

4.3 Schottky pairs and cation Frenkel pairs in metastable states

To examine the influence of the reference state on defect formation energies, we studied two metastable configurations that are (1) 1.16 eV and (2) 5.23 eV (in the unit of eV per supercell) above the "ground" state in their supercell energies. (In reality, (2) is unlikely to exist.) The same hybrid calculations were performed giving the results in **Fig. 4a-b** for (1) and **Fig. 4c-d** for (2). In each case, $E_{f,\text{Schottky}}$ is smaller than $E_{f,\text{cation Frenkel}}$ in both the lowest and the mean energies of the distribution. Comparing **Fig. 4** and **Fig. 3**, we note that the values of $E_{f,\text{Schottky}}$ and $E_{f,\text{cation Frenkel}}$ of the metastable states are smaller than their counterparts of the "ground" state. Moreover, the less stable the metastable configuration, the lower the $E_{f,\text{Schottky}}$ and $E_{f,\text{cation Frenkel}}$. Indeed, the lowest $E_{f,\text{Schottky}}$ in (2) is negative (**Fig. 4c**), indicating (2) is not even metastable if the species are allowed to exchange with the chemical reservoir to spontaneously form a Schottky pair. On the other hand, the handicap of a higher-energy metastable state is always more severe than the advantage of a lower formation energies of defects, the more so the less stable the state. (In (1), the formation energy of a Schottky pair is 0.28 eV —1.63 eV in **Fig. 4a** vs. 1.91 eV in **Fig. 3d**—lower than that in the "ground" state, but state (1) is 1.16 eV more energetic than the "ground" state. The gap is even larger between (2) and the "ground" state.) Therefore, unless the metastable state is energetically very close to the "ground" state, it is thermodynamically and kinetically unrealistic to expect them to provide Schottky pairs that contribute cation vacancies for diffusion.

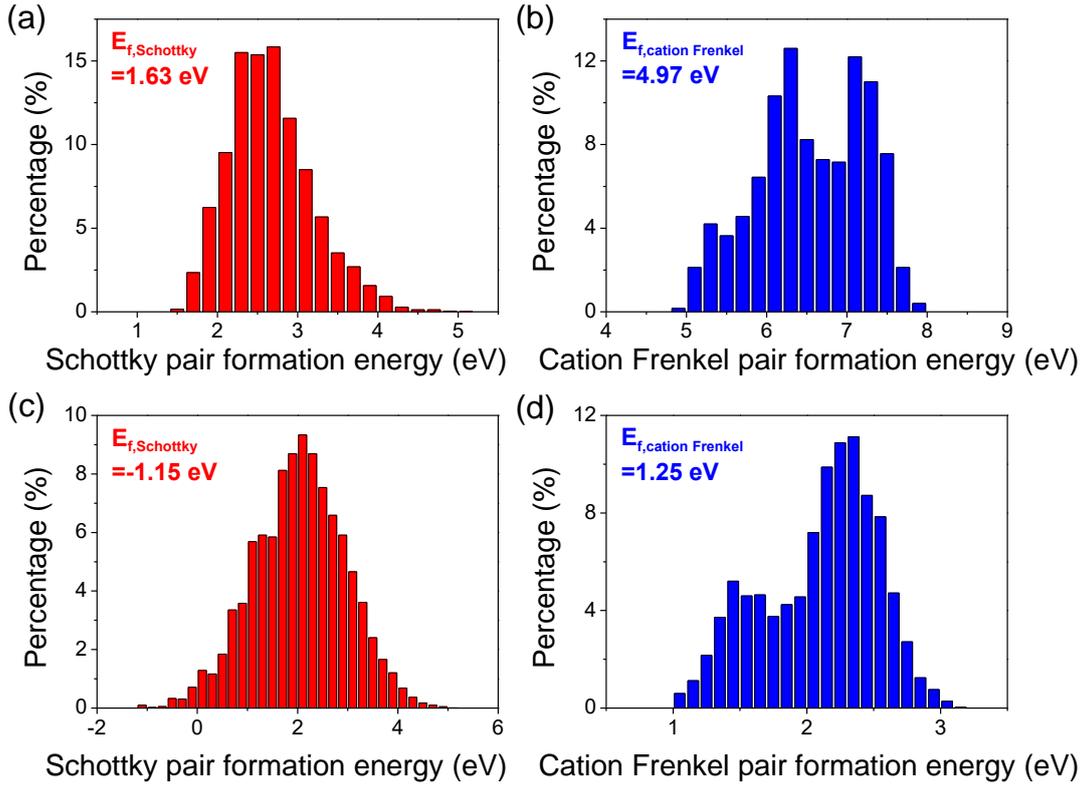

**Figure 4** Energy distributions to form a (a) Schottky pair or (b) cation Frenkel pair in a metastable YSZ having 1.16 eV higher supercell energy than the "ground" state; likewise for forming a (c) Schottky pair or (d) cation Frenkel pair in a metastable YSZ having 5.23 eV higher supercell energy. Their assemble averages between 1400-1700 K are listed at the upper-left corners. Data from *ab initio* calculations.

This study of metastable states reaffirms our conclusion in Section III that Schottky pairs always dominate over cation Frenkel pairs and cation vacancy is the dominant cation defect. Moreover, since the search for the lowest energy state in Paper I was not exhaustive and the *de facto* "ground" state we used may not be the true ground state, we should regard the $E_{f,\text{Schottky}}$ value in **Fig. 3d**, 1.91 eV, as a lower bound. (The *de facto* ground state is statistically valid up to a sample size of ~$10^9$ random configurations, which is still small compared to the $10^{42}$ configurations available in a 3×3×3 supercell; see Section VI of Paper I.) However, we expect the formation energy difference to be less than the energy difference between the true ground state and the *de facto* ground state if we can extrapolate the comparison between **Fig. 4a** and **Fig. 3d**: The difference of 0.28 eV in formation energy is much less than the difference of 1.16 eV in the supercell state energy.

4.4 Local structure of cation vacancies

Since least energetic vacancies are likely to be predominant, it is instructive to interrogate the correlation between $E_{V_{Zr}}$ (and $E_{V_Y}$) and the local structure within the framework of packing preferences established in Paper I. For this purpose, we studied the 10 least energetic $V_{Zr}$, and correlated their *ab initio* supercell energies $E_{V_{Zr}}$ with supercell's bond valence energy (**Fig. 5a**, blue bars) and electrostatic energy (**Fig. 5b**, blue bars). (See Part I on the method of the required calculations. Here, the electrostatic energy is obtained by treating the vacancy-containing supercell as having a set of point charges made of $Zr^{4+}$, $Y^{3+}$ and $O^{2-}$, in addition to a negative charge of, say –4 if the supercell contains a $V_{Zr}$, and surrounding the supercell by a uniform background charge of +4 in total.) Since these vacancies are likely to form by removing the most energetic $Zr_{Zr}$ in the ground state, we also studied how the vacancy-containing supercell energy correlates with Zr's environment—its numbers of O-1$^{st}$ nearest neighbors, denoted as O-1NN (**Fig. 5c**), and 2$^{nd}$ nearest neighbors, O-2NN (**Fig. 5d**).

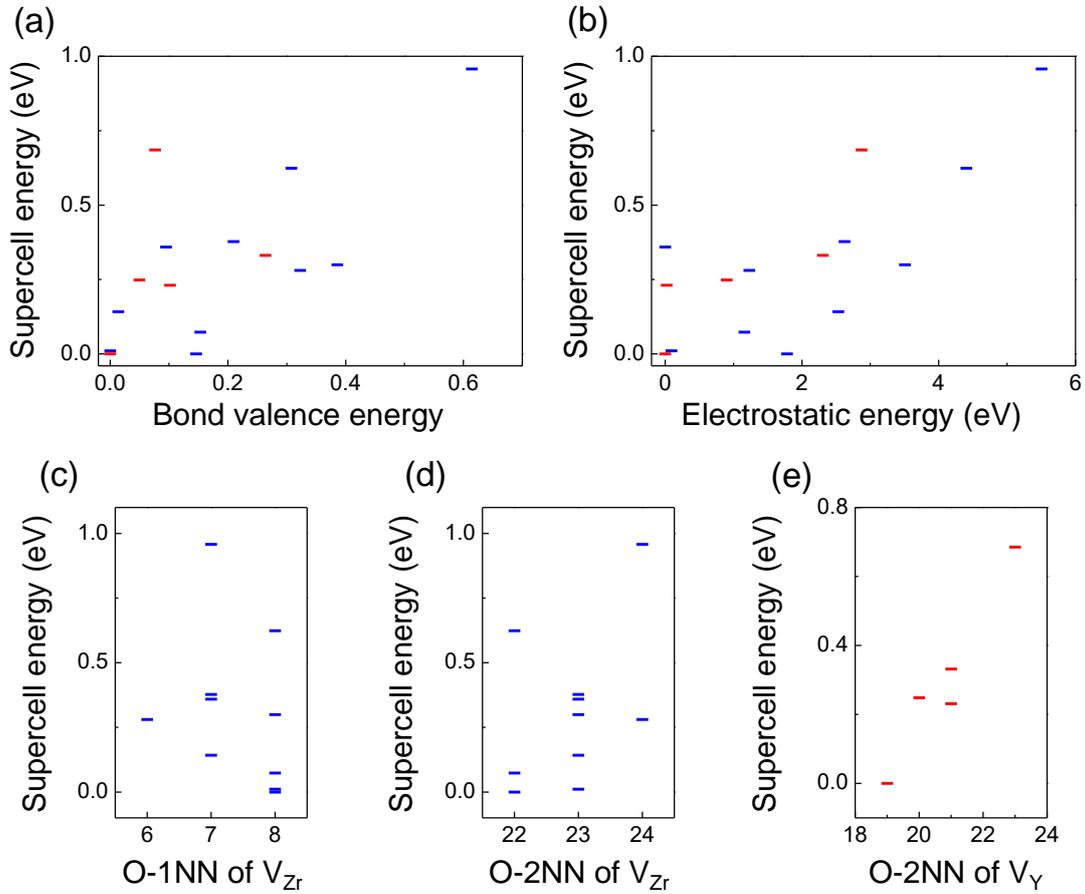

**Figure 5** (a) Bond valence energy and (b) electrostatic energy correlates with the energy of supercell that contains one $V_{Zr}$ (blue bars) or $V_Y$ (red bars). The lowest such energies are used as the energy references (i.e., zero energy) for comparison within the same set, blue or red. Also compared with the supercell energy are (c) O-1NN and (d) O-2NN of $V_{Zr}$, and (e) O-2NN of $V_Y$. The $V_Y$ in this set all have 8 O-1NN.

Because of the small sample size and the considerable structural heterogeneity of different defects in the sample, correlations were relatively weak but still enough to recognize the following trend. A less energetic $V_{Zr}$ is in a supercell of a lower bond valence energy and electrostatic energy, and it originates from a Zr site that has an unfavorably high O-1NN number (fewer oxygens are preferred, see Paper I) and unfavorably low O-2NN number (more oxygens are preferred). According to Paper I, the cation-O-1NN preferences mainly originate from size consideration (Zr being undersized relative to Y) while the cation-O-2NN preferences mainly from electrostatic charge consideration ($Y^{3+}$ being -1 relative to $Zr^{4+}$). Since $V_{Zr}$ is highly negative, it introduces another charge consideration that favors placing positively charged $V_O$ near $V_{Zr}$, i.e., $V_{Zr}$ originating from a Zr site of fewer O-1NN and O-2NN. This new consideration does not contradict the packing rule established in Paper I for O-2NN, but it does contradict the size consideration of Paper I for O-1NN. Yet the correlation in **Fig. 5c-d** is fully consistent with the rules of Paper I. Therefore, the additional charge consideration for $V_{Zr}$ is not as important as the size-dominated packing consideration for the O-1NN environment.

We similarly examined the 5 least energetic $V_Y$, their $E_{V_Y}$, bond valence (**Fig. 5a**, red bars) and electrostatic energies (**Fig. 5b**, red bars), and again found some correlations. Regarding oxygen coordination, we note that our "ground" state only contains Y that has 8 O-1NN, which conforms to Y's packing preference—dictated by the size consideration for 1NN. Regarding O-2NN, we shall follow two charge considerations. (a) According to Paper I, $Y^{3+}$ being -1 relative to $Zr^{4+}$ prefers more $V_O$, hence fewer O-2NN; conversely, the least energetic $V_Y$ should originate from a Y site with the highest O-2NN. (b) $V_Y$ being highly negative -3 prefers more $V_O$, hence fewer O-2NN; conversely, the least energetic $V_Y$ should originate from a Y site with the lowest O-2NN. Apparently, (b), which is a new consideration, overrides (a), which came from Paper I, so that the trend in **Fig. 5e** is manifest. (**Fig. 5e** has less scattered data than **Fig. 5c-d** since Y always has 8 O-1NN.)

In summary, as in Paper I, we find for cation vacancies that their O-1NN preferences mainly originate from size consideration while their O-2NN preferences mainly from electrostatic charge consideration. The majority $V_{Zr}$ prefers to have $V_O$ as 2NN, not 1NN; the minority $V_Y$ prefers to have $V_O$ as both 1NN and 2NN.

**V. Migration barrier in YSZ**

5.1 Migration barrier in the "*ground*" state

To calculate the migration barrier for a cation vacancy in YSZ, we again considered the multiple possibilities of where migration may begin and end in a supercell, which is not invariant in translation. To keep the computational task tractable, we started with the least energetic $V_{Zr}$ created at the most energetically unfavorable Zr site, termed *A*, and find the barriers for Zr to enter *A* from 6 out of the 12 neighboring sites. After two neighboring Zr with the lowest barriers left their sites, *B* and *C*, we again calculated 6 migration paths for Zr to enter *B* or *C* following the same scheme. (Having Zr returning from *A* to *B* or *C* was one of the 6 paths.) The energy profiles (one shown in **Fig. 6**) along the above paths in a 3×3×3 supercell were obtained using the *ab initio*-NEB method, and their migration barriers in the forward and backward directions are listed in **Table II**. (The average of the forward and barrier barriers is designated as the "equivalent" migration barrier of the path.)

The barriers in two directions differ because of lack of symmetry in YSZ. For example, since *A* is the most stable state, the backward barrier to return to *A* is always lower than the forward barrier in **Table II**. The barriers from *A* vary from 1.78 to 4.43 eV with the average at 2.92±0.58 eV. (The forward average is 3.22±0.70 eV, the backward 2.62±0.56 eV.)

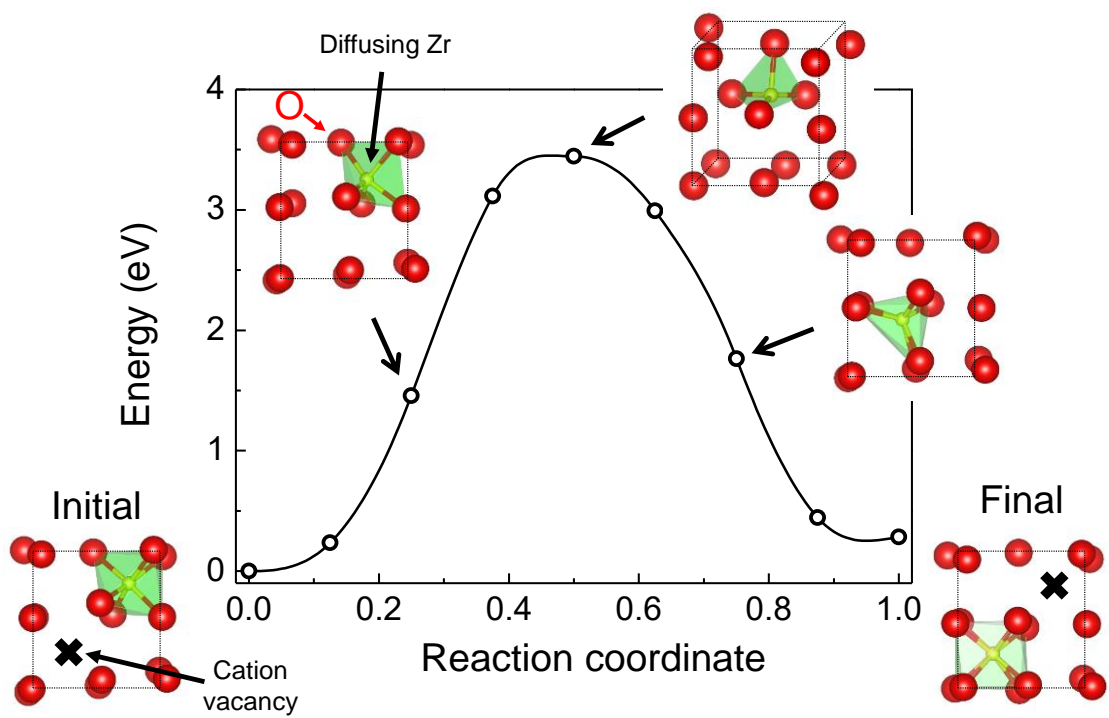

**Figure 6** Cation migration energetics and local atomic arrangements at several intermediate states.

**Table II** Migration barriers in 8YSZ for six forward /backward paths each from *A*, *B*, and *C*. The equivalent barrier (*E*) is the average of forward (*F*) and backward (*B*) barriers. The locations of $V_O$ as indexed in Fig. 9 are listed under $P(V_O)$. *Same as 4 from *A*, #Same as 5 from *A*.

| # | Hopping from *A* | | | | Hopping from *B* | | | | Hopping from *C* | | | |
|---|---|---|---|---|---|---|---|---|---|---|---|---|
| | *F* (eV) | *B* (eV) | *E* (eV) | $P(V_O)$ | *F* (eV) | *B* (eV) | *E* (eV) | $P(V_O)$ | *F* (eV) | *B* (eV) | *E* (eV) | $P(V_O)$ |
| 1 | 3.36 | 3.30 | 3.33 | 15 | 3.42 | 2.54 | 2.98 | None | 2.84 | 3.26 | 3.05 | 11 |
| 2 | 4.43 | 3.01 | 3.72 | 11 | 4.16 | 3.19 | 3.68 | None | 3.37 | 2.79 | 3.08 | 12 |
| 3 | 3.16 | 2.76 | 2.96 | 17 | 4.18 | 3.41 | 3.80 | 6, 17 | 2.67 | 2.75 | 2.71 | 1 |
| 4 | 3.01 | 2.73 | 2.87 | 17 | 2.73* | 3.01* | 2.87* | 17* | 4.34 | 3.73 | 4.04 | None |
| 5 | 2.26 | 1.78 | 2.02 | 1 | 3.14 | 2.81 | 2.98 | 3 | 1.78# | 2.26# | 2.02# | 1# |
| 6 | 3.10 | 2.15 | 2.63 | 11 | 3.45 | 3.17 | 3.31 | 3, 12 | 3.88 | 2.81 | 3.35 | None |

5.2 Migration kinematics

A glimpse of migration kinematics is afforded by snapshots taken along the path. One set of snapshots is shown as insets in **Fig. 6**, and another in **Fig. S1**. Here, the migrating Zr is in yellow, and the $V_{Zr}$ with which it exchanges locations is marked by a cross. The kinematics have the following features. (a) Except for the migrating Zr and $V_{Zr}$, other ions only experience minor local relaxations. (b) Migration is not along a straight line from the initial site to the final site (its midpoint marked as M in **Fig. 7**). Instead, the saddle point veers toward the vacant cation interstitial site (marked as I in **Fig. 7**). (c) While two nominally equivalent cation interstitial sites are marked in **Fig. 7**, long range interactions originating from ions outside the ones in **Fig. 7** led the NEB-calculated path to prefer one site over the other. (d) The saddle-point Zr has fewer oxygen neighbors than the initial/final-site Zr, 4 oxygens (3, 8, 9 and 10, as marked in **Fig. 7**) in **Fig. 6** and 5 oxygens (3, 4, 8, 9 and 10) in **Fig. S1** in the former compared to 8 oxygens in the latter. (e) To compensate for the lower coordination number, the saddle-point Zr-O polyhedron has shorter (1.9-2.0 Å between saddle-point Zr and oxygen 9/10 vs. normally ~2.2 Å) hence stronger Zr-O bonds than the ones in a normal 7- or 8-coordinated polyhedron. (f) To make room for Zr passage, oxygen 9 and 10 are pushed out. (g) If we construct a triangular prism of oxygens {3, 4, 7, 8, 9, 10}, then the migration path enters the prism from one side 3-4-10-9 and leaves from the other side 7-8-10-9.

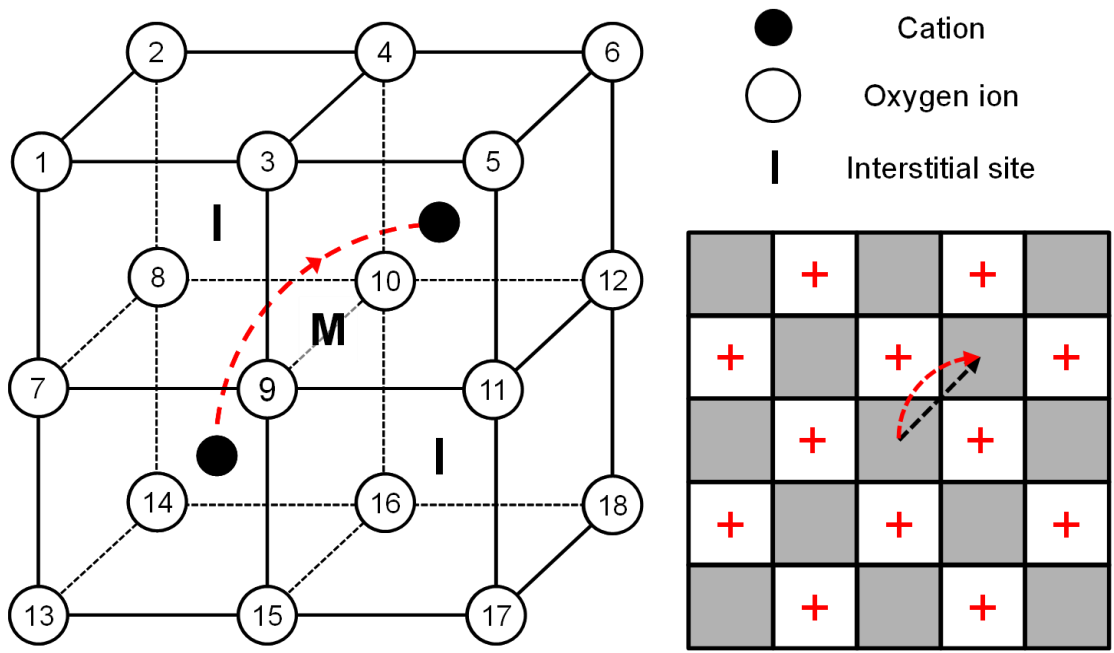

**Figure 7** (Left) Schematic of 1×1×½ slab with one Zr and one $V_{Zr}$ exchanging sites along the red dotted path. Oxygen sites indexed from 1 to 18, two interstitial locations marked as I, and the center of the slab marked as M. (Right) Schematic of cation plane showing a checkerboard arrangement of grey $ZrO_8$ polyhedra. Cation can either migrate via M (black dash line) or on a detour via an interstitial site (red dash curve).

5.3 Charge transfer and long-range disturbance

Closely related to the coordination change is charge transfer during migration. Since our calculations did not allow redox reactions, any charge transfer must be caused by either ion motion that drags the surrounding electron cloud or electron redistribution between neighboring ions, and it should be mostly reversible once cation migration has passed. In crystal chemistry terms, this corresponds to a "bond-valence" change due to the changing coordination number and bond distance.

In the example shown in **Fig. 8**, the *ab initio*-calculated charge density difference between the saddle-point state and the initial state is plotted in **Fig. 8a**, and similarly, the difference between the final state and the saddle-point state is plotted in **Fig. 8b**. Here, the migrating Zr is marked in purple, electron surplus is represented in blue, and electron deficit is in yellow. As Zr migrates toward the saddle point in **Fig. 8a**, it attracts oxygen O2, O3 and O4 and draw in more bonding electrons. Meanwhile, another surrounding oxygen O1 is left behind and drawn away by a nearby cation, which happens to be Y in **Fig. 8a**. As Zr migrates away from the saddle point in **Fig. 8b**, it sheds the bonding electron with O2 but simultaneously draws in bonding electrons from O3, O4, O5 and O6, which are retained in the final state. Comparing **Fig. 8a** and **b**, we can see numerous oppositely pointing "dipoles", directing from yellow to blue, several shown by black arrows. Remarkably, such bonding changes have propagated across the entire length of the 3×3×3 supercell as shown in **Fig. 8c**, suggesting very slow attenuation against the drastic bonding perturbation that accompanies cation migration. This may be taken as another evidence of the compliant cation sublattice first noted by Li *et al*. [22-24], who also cited its relatively soft acoustic phonon that is consistent with YSZ's relatively low Young's modulus (200 GPa) and shear modulus (80 GPa) despite the high melting point.

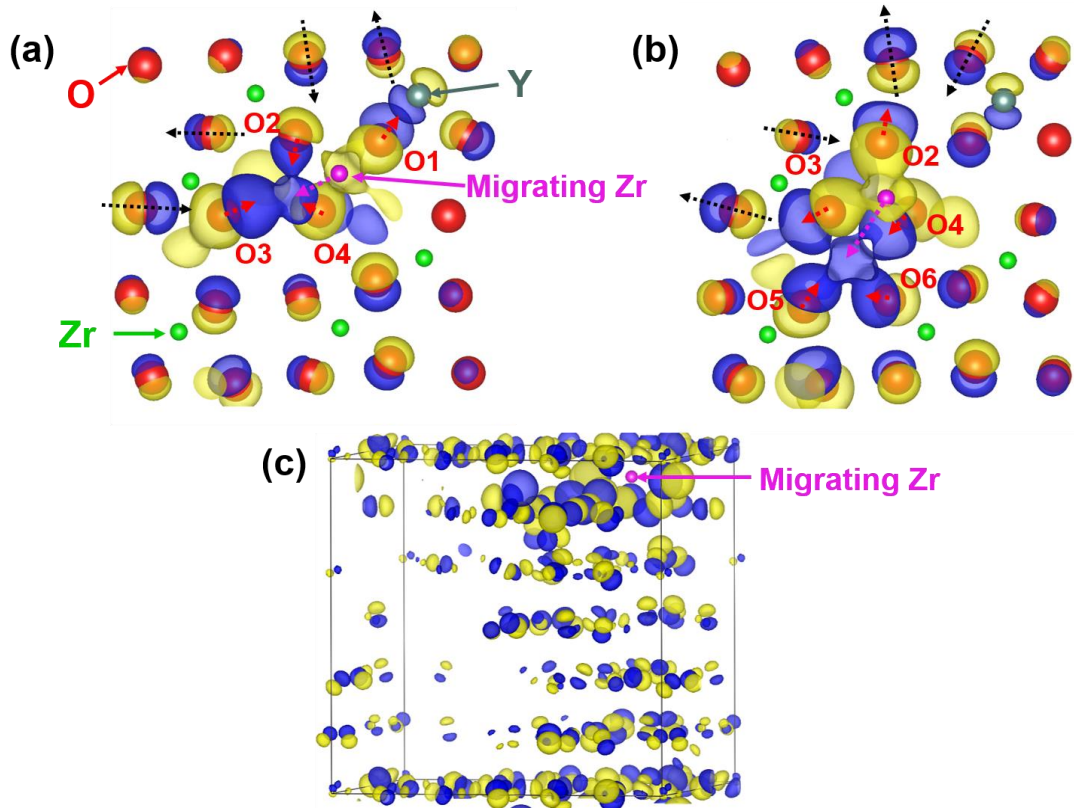

**Figure 8** Electron charge density transfer (a) from the initial state to saddle point, and (b) from saddle point to final state, drawn as surfaces of iso-charge-transfer of 0.02 electron/Bohr$^3$ around the migrating Zr. Electron surplus is represented in blue and deficit is in yellow. (c) Same as (a) shown for the entire 3×3×3 supercell at a smaller magnification. The iso-charge-transfer drawn is 0.05 electron/Bohr$^3$.

5.4 Correlating barrier height to saddle-point's local structure

We next seek to correlate barrier height with the local structure when the cation is passing through the saddle point. As shown in **Fig. 7**, ZrO$_8$ polyhedra (in grey) are arranged in a checkerboard pattern, and the empty space (in white) provides interstitial sites. To take advantage of the empty space, Zr during migration deviates from the straight-line path through M in **Fig. 7** to follow the schematic dotted curve nearby. Naively, one would expect less need for such detour if there is more space between O9 and O10 to allow a direct passage, or if O9 and O10 can move out of way to make room for Zr. The latter is indeed the case as illustrated in **Fig 9a** by the inverse correlation between $L_{Zr-M}$, the distance between M and Zr at the saddle point, and $L_{O9/O10}$, the total displacements of O9 and O10 calculated in the following way. When Zr moves from the initial point to the saddle point, O9 moves by $L_1$ and O10 by $L_2$; when Zr moves from the saddle point to the final point, O9 moves back by $L_3$ and O10 by $L_4$. The sum of these four displacements, all taken as positive numbers, is $L_{O9/O10}$.

The interplay between $L_{Zr-M}$ and $L_{O9/O10}$, expressed in $L_{Zr-O9/O10}$, which is the average bond length of Zr-O9 and Zr-O10 at the saddle point, has a decisive influence on the migration barrier as shown in **Fig. 9b** with a correlation factor of –0.76. Since a shorter bond distance implies a higher short-range repulsion, we also expect a positive correlation between the migration barrier and such repulsion. This is demonstrated by **Fig. 9c**, (correlation factor: 0.78) in which the Buckingham potential in the empirical potential [14, 19] was used to calculate the short-range repulsions between the migrating Zr and nearby oxygens up to 3 Å. It is a reasonable result since the less repulsion the neighboring ions exert on the migrating cation, the less migration barrier the migrating ion should experience. The most important differentiator that causes the total repulsion to vary is the short-range repulsion from O9 and O10, which has a correlation factor of 0.75 with the migration barrier as shown in **Fig. 9d**. This is not surprising because, at the saddle point, the Zr-O9 and Zr-O10 distances are significantly shortened, from normally ~2.2 Å before migration to 1.9-2.0 Å at the saddle point. Therefore, size consideration is of pivotal importance in determining the ease of cation migration in YSZ.

In the order of increasing importance, other weaker correlations between migration barrier and local structure parameters are with (a) $L_{Zr-M}$, the distance between Zr and M at the saddle point (correlation factor: 0.15); (b) $L_{O9/O10}$ (correlation coefficient: –0.23); (c) O9-O10 distance in the initial/final configuration (correlation coefficient: –0.25); and (d) cation size, which is related to the charge state represented by the Bader charge (correlation coefficient: 0.42) [25]. Naturally, these weaker correlations are mostly washed out by the noise (i.e., the configuration-to-configuration variation of YSZ), but on the whole all of the above are consistent with the idea that size consideration is of paramount importance: A larger Zr separation from O9 and O10 at the saddle point, a longer O9-O10 separation, more compliant O9 and O10, and a smaller cation size all help lower the barrier. This is because it is the misfit between the migrating cation size and the statically or dynamically available space between O9 and O10 that determines the short-range repulsions, which further extends to influences felt at much longer distances via slowly converged elastic and electrostatic interactions as indicated in **Fig. 8c**.

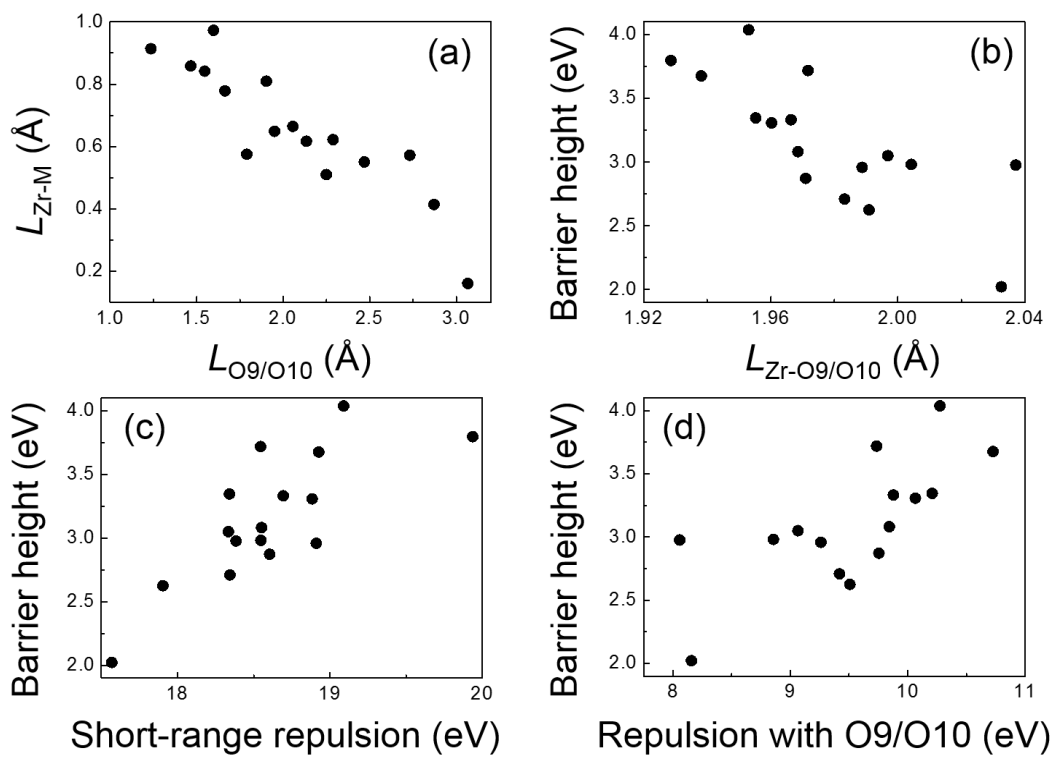

**Figure 9** Correlations between static and dynamic local structure at saddle point and migration energetics. (a) $L_{Zr-M}$ vs. $L_{O9/O10}$, (b) migration barrier from *ab initio*-NEB calculations vs. $L_{Zr-O9/O10}$, (c) same barrier vs. short-range repulsions between migrating Zr and neighboring oxygens up to 3.0 Å, and (d) same barrier vs. short-range repulsions between migrating Zr and O9/O10. $L_{Zr-M}$: the distance between M and Zr at the saddle point; $L_{O9/O10}$: the total displacements of O9 and O10 during the migration process (see text for more detail); $L_{Zr-O9/O10}$: the average bond length of Zr-O9 and Zr-O10 at the saddle point.

5.5 Y Migration

Similar *ab initio*-NEB calculations were performed for Y migration. Three migration paths were examined and they share the same migration kinematics as described above for Zr migration. The equivalent diffusion barriers are 3.28±0.77 eV (forward: 3.16±0.64 eV; backward: 3.39±0.90 eV.) They are slightly higher than those for Zr, but the data scatter is large and we only examined 3 paths.

Since a cation vacancy in YSZ can be used for both Zr and Y diffusion (thus the same defect formation energy), any diffusivity difference between the two must come from different migration barriers. According to tracer diffusion data, Y diffusion is 2.5-4.3× that of Zr between 1300 $^{o}$C and 1676 $^{o}$C [5, 26], which corresponds to a lower migration barrier by ~0.2 eV in Y. On the other hand, such kinetic advantage should have led to Y segregation at the cathode in electro-diffusion, yet $Y^{3+}$ actually segregates at the anode instead [27, 28]. The latter observation places an upper bound for $Y^{3+}$ diffusivity at 4/3× that of $Zr^{4+}$, suggesting a <0.1 eV difference in migration barrier. Clearly, for a complicated structure like YSZ, these differences in migration barrier are too small to be resolved by *ab initio* calculations employed here. So we did not pursue it further.

## VI. Comparison with experiments and past calculations

6.1 Diffusion as a random walk process

Atomic migration is a multi-step random-walk process that must reach a long enough distance to become effective, or else the vacancy and the atom may return to their original sites. A random walk in YSZ sees myriad local structures and barrier heights, which vary from step to step and from path to path. With many paths available, the preferred one is the easiest one, and on this easiest path it is the highest barrier that is rate controlling. This random-walk problem is similar to the conduction problem of a network circuit that has both serial and parallel elements. In the following, we will apply this idea to evaluate the rate-controlling barrier for Zr.

To proceed, we first postulate that three independent migration steps are needed for Zr to "permanently" escape its original site and not look back. This is a reasonable assumption: It can be shown that, while after two steps the probability for the vacancy to return to the original site is still 1/12, after three steps it drops to 1/36, which seems quite small. (Here, we use the fact that each $V_{Zr}$ has 12 nearest Zr neighbors.) We thus calculated the time required to statistically sample all available three-step paths in the following way. (a) Referring to **Table II**, we consider taking one step each from *A*, *B* and *C* to constitute a three-step path, thus obtaining $6^3=216$ such paths (6 available steps each from *A*, *B* and *C*). (b) Consider a three-step migration as a serial process, we see the time $\tau_m$ required to go through it to be proportional to $\exp(E_{m,i}/k_BT)+\exp(E_{m,j}/k_BT)+\exp(E_{m,l}/k_BT)$, where $E_{m,k}$ is the barrier height of step *k* (*k=i, j, l*) along the path *m* and we assume the correlation factor and attempt frequency are the same for each jump. (c) Since the probability of taking path *m* to "escape" within a unit time is $\tau_m^{-1}$, the total escape probability per unit time when having *M* such parallel paths available (*M*=216 in our case) is $\sum_{m=1}^{M} 1/\tau_m$. Therefore, the ensemble average of the migration barrier for the random walk can be obtained from the slope of the

$$\ln \sum_{i,j,l=1}^{6} 1/\left[\exp\left(E_{m,i}/k_BT\right)+\exp\left(E_{m,j}/k_BT\right)+\exp\left(E_{m,l}/k_BT\right)\right]$$ vs. 1/*T* plot, where *i* is

a path from *A*, *j* a path from *B* and *l* a path from *C*, providing 216 three-step paths. Over 1,400-1,700K, the plot gives a migration barrier of 2.88 eV to 2.97 eV depending on whether the equivalent or forward barriers in **Table II** are used.

6.2 Experimental data and past calculations

Diffusivity *D* may be theoretically estimated using its Arrhenius form

$$D = D_0 \exp(-\frac{E_a}{k_B T}) = \frac{fa^2 v}{[V_O]^2} \exp(-\frac{E_f + E_m}{k_B T}) \quad (9)$$

Here $D_0$ is a pre-exponent factor, $k_B$ is the Boltzmann constant, $T$ is the absolute temperature, $E_a$ is the activation energy, $f$ is the correlation factor, $a$ is the jump distance, $v$ is the attempt frequency, $[V_O]$ is the fraction of vacant sites on the oxygen sublattice, and $E_f$ and $E_m$ are the defect formation and migration energy, respectively. (The law of mass action applied to the Schottky defect formation reaction gives $\frac{1}{[V_O]^2} \exp(-\frac{E_f}{k_B T})$ as the fraction of vacant cation sites on the cation sublattice.) For cation diffusion in YSZ, $f=0.783$ for the face-centered-cubic cation sublattice, $a=3.64$ Å, $v$ is taken to be $10^{12}$-$10^{14}$ s$^{-1}$, $[V_O]=0.074$, and $E_f$ and $E_m$ are the calculated enthalpies, $E_f=1.91$ eV from the Schottky defect in Section IV and $E_m=2.97$ eV in Section VI, giving an activation energy of 4.88 eV, compared very well to the activation energy of 4.5-6.1 eV from experimental data [1-5]. For $v=10^{13}$ s$^{-1}$, our theoretical estimates shown in red in **Fig. 10** are in very good agreement with the experimental data (in black) of cubic zirconia [1-3, 5] of comparable compositions to 8YSZ over quite a wide range of temperature.

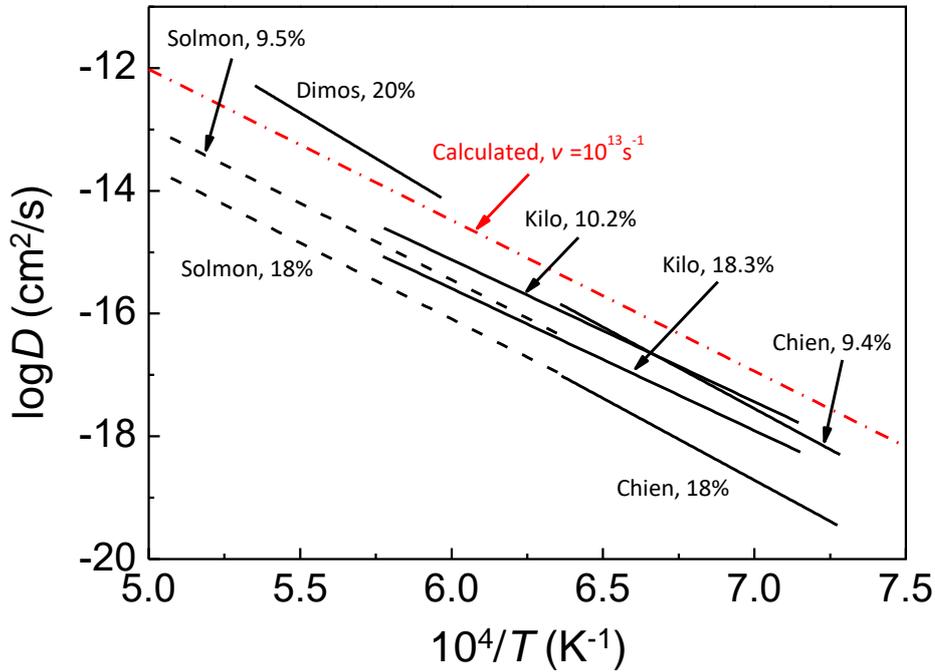

**Figure 10** Predicted cation diffusivity of this study (red) in agreement with experimental data from tracer diffusion (Solmon's 9.5YSZ and 18YSZ [5], Kilo's 10.2YSZ and 18.3YSZ [1]), creep (Dimos's 20YSZ [3]) and shrinkage of dislocation loop (Chien's 9.4YSZ and 18YSZ [2]).

As mentioned in the Introduction, all previous theoretical and computational studies, which used empirical potentials only, have consistently found an activation energy >10 eV, of which 6.5 eV or more is for forming a cation vacancy and 5 eV for it to migrate. (See **Table III** for a summary of previous simulation results.) As also mentioned in Part I, there is a systematic overestimate of the supercell energy by the empirical potential calculations because of their overestimate of the electrostatic energy. To interrogate whether this is also the case here, we adopted the structures of the saddle point and the initial/final state obtained by our *ab initio*-NEB calculations, then used the empirical potential to calculate the supercell energies of these states, and finally took their difference as the migration barrier. The results are compared with the *ab initio*-NEB barrier calculations in **Fig. 11** for both the forward (blue) and equivalent (red) migration barriers. As before, we find a positive >2× correlation: The barrier calculated using the empirical potentials is more than twice as high as from *ab-initio* method. Since the same structure is used in the two calculations, the difference can only come from the computational method itself. Therefore, without doubt, the empirical potential calculations that overestimate the electrostatic interaction gave the wrong activation energies.

**Table III** Summary for calculated cation formation and migration energies in zirconia literature. *Potential modulated to fit oxygen diffusion data.

| Method | Composition | Formation energy (eV) | Migration energy (eV) |
|---|---|---|---|
| Empirical potential [9] | *c*-ZrO$_2$ | Schottky: 11.6<br>Cation Frenkel: 24.4 | Zr: 8.5 |
| Empirical potential [10] | *c*-ZrO$_2$ | Schottky: 5.91<br>Cation Frenkel: 20.15 | / |
| | *t*-ZrO$_2$ | Schottky: 6.55<br>Cation Frenkel: 17.63 | / |
| Empirical potential | *t*-ZrO$_2$ | Schottky: 6.52 | Zr: 5.07 |

| | [11] | | Cation Frenkel: 20.44 | Y: 4.60 |
| --- | --- | --- | --- | --- |
| Empirical potential, molecular dynamics [13] | 11YSZ | / | | Zr: 4.8 Y: 4.7 |
| Modulated potential*, molecular dynamics [15] | 8YSZ | / | | Zr: 2.58 Y: 2.61 |

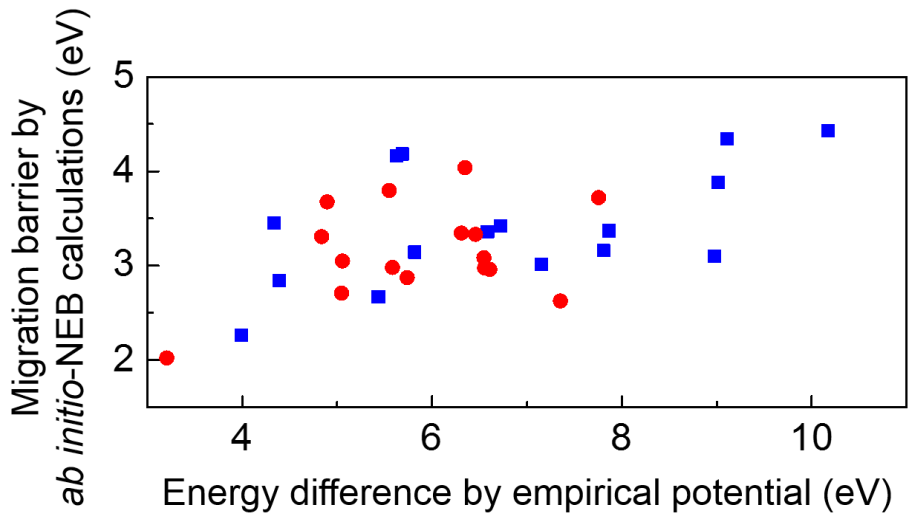

**Figure 11** For the same set of initial, final and saddle-point atomic structures, the forward (blue)/equivalent (red) migration barriers from *ab initio* calculations are smaller than those calculated by using the empirical potential.

## VII. Discussion

### 7.1 Defect formation energy

While only the formation energies of the Schottky and Frenkel pairs have been calculated in this work, one can also consider the formation energies of individual charged defects [29, 30]. For example, $E_{f,V_{Zr}}$ for V$_{Zr}$ and $E_{f,V_O}$ for V$_O$ are defined as

$$E_{f,V_{Zr}} = E_{V_{Zr}} - E_{ref} + \mu_{Zr} - 4\mu_e \quad (10)$$

$$E_{f,V_O} = E_{V_O} - E_{ref} + \mu_O + 2\mu_e \quad (11)$$

where $\mu_{Zr}$ and $\mu_O$ denote the chemical potentials of Zr and O, respectively, $\mu_e$ is the chemical potential (the Fermi level) of electrons, and $\mu_{Zr} + 2\mu_O = \mu_{ZrO_2}$ under local equilibrium conditions. Combining Eq. (10-11), we obtain the formation energy of a Schottky pair, $E_{f,Schottky} = E_{f,V_{Zr}} + 2E_{f,V_O}$, and recover Eq. (4). Using these relations, we can also use the literature data of $E_{f,V_{Zr}}$, $E_{f,V_O}$ and other interstitial defects in monoclinic zirconia [30] to calculate the pair formation energies: A Schottky pair needs 6.95 eV, a cation Frenkel pair needs 9.45 eV, and an anion Frenkel pair needs 4.11 eV. These values are similar to the results listed in **Table I** within the error expected for *ab initio* calculations.

For simplicity, our calculation has assumed fully ionized defects. While this is empirically justified by the fact that YSZ is difficult to reduce or oxidize, it contains only minimal amount of electrons and holes [31] and it is an ionic conductor, it is also justified by referring to the energy diagrams of cubic Y$_2$O$_3$ and monoclinic ZrO$_2$ in **Fig. 12** [29, 30]. For an isolated insulator, which corresponds to normal processing conditions of oxide ceramics, we may set the Fermi level at about halfway in the band gap. Indeed, all the defects are fully ionized under this condition. The Fermi-level window within which the above assumption holds is about 1.5 eV, or 2.2 eV if we ignore $O_i$ that is unimportant in YSZ. This value is comparable to the commonly used electrochemical window, 2.1 V, the Nernst potential that causes $ZrO_2$ to reduce to Zr.

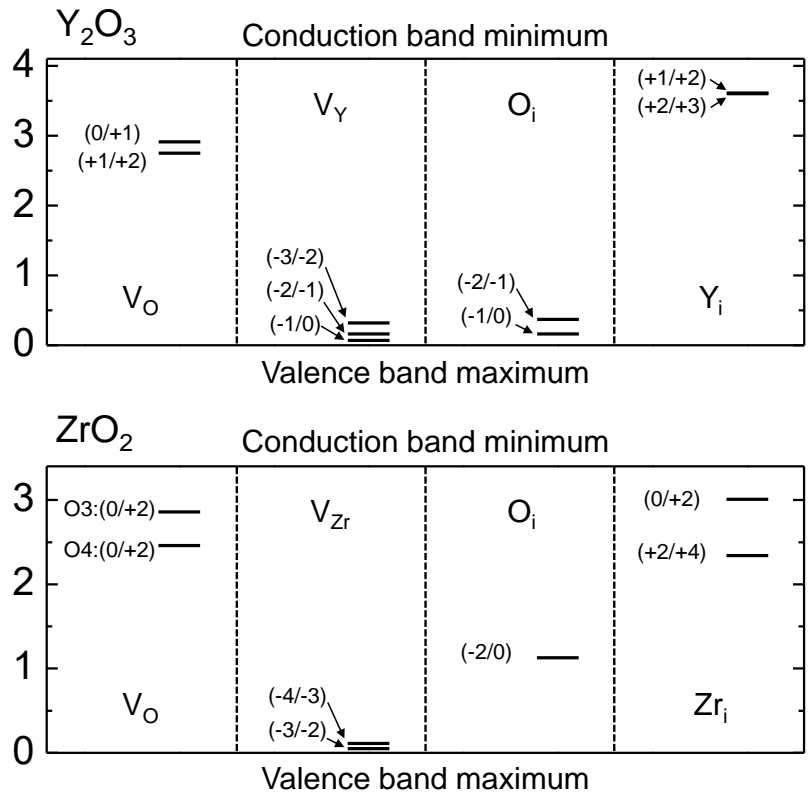

**Figure 12** Energy levels of the charge state transitions of ionic defects in $Y_2O_3$ (upper panel) and monoclinic $ZrO_2$ (lower panel). (Data from [29, 30].) In monoclinic $ZrO_2$, the O3 type of oxygen ions is coordinated with three Zr and the O4 type with four Zr. Oxygen defects in monoclinic $ZrO_2$ show negative-$U$ behavior, so their charge states disproportionate and change by 2.

7.2 The role of oxygen/oxygen vacancy

Paper I pointed out that 8YSZ with its complicated structure has a glassy energy landscape, so the true ground state cannot be reached without long-range relaxation of both cations and anions. Full relaxation is difficult because the Zr-O and Y-O environments are strongly intercorrelated, as evidenced by the distinct oxygen (vacancy) population oscillations around Zr and Y that are long-range (at least up to 5$^{th}$ NN) but out-of-phase with each other. Interestingly, as illustrated by **Fig. 8**, a single exchange between Zr and V$_{Zr}$ already entails long-range relaxations involving essentially all the ions (about 300) in the supercell. Thus, local atomic hopping is additionally coupled to another set of long-range structural relaxations. Therefore, full relaxation is certainly impossible once cation diffusion is frozen at below about 1150°C. Nevertheless, partial relaxation via much faster oxygen rearrangement may persist: It is responsible for the apparent transition in oxygen diffusivity at about 500°C [32].

We believe oxygen diffusion with a very low activation energy of about 0.5 eV above 1000°C [32] and a very high V$_O$ population plays an important role in cation diffusion, which has a migration barrier of about 3 eV. This is because the cation/cation-vacancy pair must spend much time together before it executes a successful exchange, and during the waiting period its local structure will most certainly relax, by V$_O$ diffusion, to attain a lower overall system energy. (In contrast, V$_O$-mediated relaxation of the saddle-point state is unlikely because the jump is a very brief event.) In **Table II**, the forward jumps from *A* are from the "ground" state, so no V$_O$-induced relaxation is needed. But the backward jumps to *A* and all other jumps to and from *B* and *C* must start from higher energy states. With V$_O$ repositioning, these states will likely assume a configuration that more resembles the "ground" state.

For the above reasons, the forward jumps from *A* in **Table II** are more representative of the reality and thus deserve special attention. They all start from the identical initial state, with $V_{Zr}$ surrounded by 8 oxygens without any $V_O$. Referring to **Table II**, we see the local environment within the 1×1×½ supercell of **Fig. 6,** which surrounds the $V_{Zr}$ and the migrating Zr, can be described by the site(s) of $V_O$: The $V_O$ at {11 or 15, which are equivalent sites} is the 1NN of the migrating Zr and the 2NN of the starting $V_{Zr}$; the VO at {1 or 17, which are equivalent} is the 2NN of both the $V_{Zr}$ and the migrating Zr. Note that the 3 paths involving one $V_O$ at 11/15 are all from $V_{Zr}O_8$ to $ZrO_7$, among them two paths having an identical $V_O$ at 11. Yet their migration barriers differ by more than 1 eV. This indicates that substantial heterogeneity must have arisen from the different long-range interactions due to the different environments outside the 1×1×½ slab. The 3 paths involving a $V_O$ at 1/17 are all from $V_{Zr}O_8$ to $ZrO_8$, among them two paths having an identical $V_O$ at 17. Their migration barriers, though again different, are relatively lower than those of the other 3 paths. This suggests a 2NN $V_O$ of both the $V_{Zr}$ and the migrating Zr may facilitate migration. Indeed, in addition to the paths that originate from *A*, all other paths listed in **Table II** that contain a single $V_O$ at 1/17 also have a relatively low migration barrier. On the other hand, 3 of the 4 paths that contain no $V_O$ at all have very high migration barriers (3.88-4.34 eV). Returning to the paths from *A*, besides the 6 paths listed in **Table II**, 4 out of the remaining 6 paths contain no $V_O$ and are likely to have a high migration barrier. This is not unreasonable because, to be representative of the composition of 8YSZ, a typical 1×1×½ slab must have 2 Zr and 4 O. Thus on average it can only contain 2/3 of an oxygen vacancy, i.e., 1/3 of the slabs must have no $V_O$. In this way, we have analyzed 10 of the 12 paths from *A* and confirmed the importance of $V_O$ to their migration barrier.

These observations provide the following picture of the most energetically favorable course of cation diffusion. It involves an exchange between a $V_{Zr}$ in $V_{Zr}O_8$ (the most stable $V_{Zr}$ configuration) and a Zr in $ZrO_8$ (the most unstable Zr configuration.) It also involves an environment with a 2NN $V_O$ at {1/2/17/18}. This is because, according to crystal chemistry preferences of **Fig. 3** and **Paper I**, a 2NN $V_O$ further stabilizes $V_{Zr}$ and destabilizes Zr. If the $V_O$ is instead relocated to be next to Zr, providing $ZrO_7$, it stabilizes Zr making it more reluctant to jump. Alternatively, if the $V_O$ is relocated to be next to $V_{Zr}$ (**Fig. 5c**), then it so destabilizes $V_{Zr}$ that such configuration is not seen in the "ground" state at all. More broadly speaking, to meaningfully contribute to migration, the starting configurations must be relatively low in energy, or else they will not be thermodynamically accessible, thus irrelevant (even though they may have a low migration barrier). Such a low-energy starting configuration, state *A*, is provided above. Interestingly and importantly, this state is also endowed with a low migration barrier because of the opposite crystal chemistry preferences of $V_{Zr}$ and Zr.

On a closer look, we found the 3 paths involving a $V_O$ at 1/17 from $V_{Zr}O_8$ (i.e., the ground state *A*) to $ZrO_8$ are not equivalent because the migrating path (red broken curve in **Fig. 7**) veers toward 1, not 17. If the electrostatic repulsion between positively charged $V_O$ and the migrating Zr is the determining factor, then the migrating Zr going closer to $V_O$ at 1 should experience a higher barrier, which is opposite to what we found in calculation. (Other data in **Table II** follow the same trend: $V_O$ at {3,4,7,8} is more repelling than {11,12,15,16}, yet paths with these $V_O$ have lower barriers.) So charge may not be the most important consideration. On the other hand, the observation of higher barriers when no $V_O$ is present suggests the importance of structure openness/softness, which can facilitate neighboring oxygens to rearrange and to bond the saddle-point Zr. This explains why $V_O$ at {1/17} can lower the migration barrier: $V_O$ at {1,2} allows oxygen {3,4,7,8} to more easily move around to facilitate Zr movement; likewise, an $V_O$ at {17,18} makes Zr migration easier since oxygen {9,10} can move out of the way toward site {17,18}.

The above discussion of individual case studies of oxygen vacancy and crystal chemistry preferences reaches the same conclusion as from the statistical analysis of **Fig. 9**: What is important for lowering the migration barrier is the structure openness/softness that can facilitate neighboring oxygens to rearrange and to allow cation passage, which is fundamentally a size consideration. This oxygen openness/softness needs to involve both 1NN and 2NN in order to satisfy the crystal chemistry requirement of cation-anion coordination, which is again dominated by the size consideration. In the following, we will seek to extend this conclusion to anion migration

7.3 Cation vs anion diffusion

It is remarkable that in 8YSZ the activation energy of cation diffusion (4.8 eV) is almost 10× that of anion diffusion (0.5 eV at above 1,000°C, which is the appropriate temperature to compare with cation diffusion). One reason why oxygen diffusion is easier is the presence of copious $V_O$ due to Y doping, which removes the need for defect formation, hence formation energy. But it is still remarkable that the migration energy of O (~0.5 eV) is only 1/6 of Zr migration (~3 eV). This is despite a much larger size of $O^{2-}$ compared to $Zr^{4+}$ or $Y^{3+}$. Similarly, fast anion diffusion is seen in other fluorite structured compounds ($CaF_2$, $SrF_2$, $CeO_2$), so the reason must come from the structure itself.

To understand why, we used the same setting of cubic $ZrO_2$ and calculated the migration barrier of Zr and O by *ab initio*-NEB calculations. They gave 0.37 eV for O migration and 4.64 eV for Zr migration. By inspecting the local structures at the saddle point, we identified a Zr-O bond length of 1.95 Å for Zr migration and 1.98 Å for O migration. The former value is comparable to that in YSZ (1.9-2.0 Å), and is much larger than the half spacing between O9 and O10 (2.6 Å divided by 2). So direct passage of Zr between O9 and O10 is not feasible, forcing Zr to take a curved path in **Fig. 7** (O9/O10 are still displaced by 0.5 Å). In comparison, the latter value is closer to the half spacing between the two Zr (3.6 Å divided by 2). So a direct passage of O is much more feasible and only need to displace the two Zr by less than 0.2 Å. Therefore, fluorite structure is much more open for anion migration than for cation migration, which accounts for its well-known property of faster anion diffusion.

## VIII. Conclusions

(1) Defect formation energies in zirconia polymorphs and 8YSZ have been calculated using *ab initio* methods. Schottky pairs are easier to form than cation Frenkel pairs, providing cation vacancies as the dominant cation defect species. In 8YSZ, cation vacancy formation requires about 2 eV.

(2) Cations migrate by exchange with a neighboring vacancy, which encounters a barrier from 2.26 to 4.43 eV depending on the local environment. The effective barrier for a multiple-step random-walk is about 3 eV. The predicted cation diffusivity is in good agreement with the experimental data of 8YSZ, with activation energy of about 5 eV.

(3) Cation hops by making a detour via an empty cation-interstitial site. At the saddle point, the migrating cation displaces two nearest oxygen ions with much shortened Zr-O bonds, which raises the short-range repulsions and causes a long-range disturbance of the surroundings. This explains the relatively difficulty in cation diffusion. In contrast, oxygen hopping minimally disturbs neighboring cations with much longer Zr-O bond distances, which explains why cubic zirconia is a fast oxygen conductor.

(4) A cation vacancy prefers to be surrounded by 8 oxygens; some oxygen vacancy as the next nearest neighbor will further stabilize it. Its easiest exchange is with a least stable Zr neighbor—the one surrounded by 8 oxygens. The migrating Zr prefers to travel through a neighboring empty-interstitial subcell, a ½×½×½ subcell that contains seven oxygens. Lastly, the more open and compliant the oxygen environment, the easier cation migrates. These tendencies are all fundamentally rooted in size consideration.

(5) Mobile oxygen vacancies play an important role in enabling cation diffusion by providing the above preferred local environments and oxygen softness for the cation vacancy, the Zr neighbor to be exchanged, and the oxygen vacancy in the 7-oxygen subcell surrounding the saddle point.

**Acknowledgements**

This work was supported by the Department of Energy (BES grant no. DEFG02-11ER46814) and used the facilities (LRSM) supported by the U.S. National Science Foundation (grant no. DMR-1120901). JL acknowledges support by NSF DMR-1410636.